# Epitaxial Na$_x$CoO$_2$ Thin Films via Molecular-Beam Epitaxy and Topotactic Transformation: a Model System for Sodium Intercalation

S. D. Matson[1], J. Sun[1], J. Huang[1], D. J. Werder[2], D. G. Schlom[1,3,4], and A. Singer[1*]

[1]Department of Materials Science and Engineering, Cornell University, Ithaca, NY 14853
[2]Platform for the Accelerated Realization, Analysis, and Discovery of Interface Materials (PARADIM), Cornell University, Ithaca, New York 14853
[3]Kavli Institute at Cornell for Nanoscale Science, Cornell University; Ithaca, NY 14853, USA
[4]Leibniz-Institut für Kristallzüchtung; Max-Born-Straße 2, Berlin 12489, Germany
*asinger@cornell.edu

**Abstract**
Renewable energy sources such as solar and wind are critical to combatting global warming. Nevertheless, their intermittent energy generation requires the development of large-scale grid energy storage, in contrast to the on-demand generation of coal-based power plants. Sodium-ion batteries offer a promising potential technology, yet because sodium ions are larger than lithium ions, sodium-ion intercalation results in more drastic structural rearrangements. An improved understanding of structural dynamics and ionic diffusion pathways is crucial to developing more durable sodium-ion batteries. Here we synthesize epitaxial Na$_x$CoO$_2$ by using molecular-beam epitaxy and topotactic transformation. In the synthesized epitaxial films, the CoO$_2$ layers are canted with respect to the film surface, allowing electrochemical extraction of sodium ions, which we confirm via *ex-situ* x-ray diffraction. We anticipate the epitaxial thin films reported here to enable future operando studies of interfaces, subtle lattice distortions, and microstructure during electrochemical cycling.

**Introduction**

Lithium-ion batteries are ubiquitous, yet the insufficient resources of lithium, cobalt, and other transition metals prevent lithium-ion batteries from meeting all our energy demands[1,2]. Sodium-ion batteries are a promising low-cost alternative[3,4,5]: their chemistry is similar to the more mature lithium-ion battery technology, sodium-ion batteries can be made with more abundant transition metal oxides[6], and sodium is abundant itself[3]. While sodium-ion batteries are most compelling for grid storage, they could also offer viable alternatives for powering portable electronics and electric vehicles and have a stronger foothold as we steadily deplete the lithium available today[3].

For being viable as an alternative energy storage solution, sodium-ion intercalation compounds such as layered transition metal oxides have yet to show the required durability. The materials degrade rapidly, and one of the challenges preventing further development is the limited understanding of the intercalation-induced degradation mechanisms. A typical positive electrode comprises nanoparticles of the active material, carbon black for improved electron conductivity, and polymer binders for mechanical stabilization, all commonly added during cathode slurry formation[7]. Recent research successfully revealed some of the internal mechanisms occurring in the active material during intercalation, yet the multi-component nanoparticulate morphology makes it inherently challenging to study fundamental mechanisms. Much insight could arise from investigating single crystals, yet the intercalation dynamics and poor electric conductivity limit the crystal sizes that can be efficiently intercalated to the microscale. Epitaxial thin films combine the nanometer length scale for facile intercalation of ions perpendicular to the film surface and long-range order over millimeters parallel to the surface to acquire high-quality, single crystal-like operando x-ray data, inaccessible in common nanoparticulate systems. The films would also enable operando optical microscopy of intercalation dynamics in real-time at the micrometer level and operando access to electronic transport.

Like $Li_xCoO_2$, a common commercial cathode material, $Na_xCoO_2$ can reversibly store sodium ions, making it a potential cathode material[8]. Notably, $Na_xCoO_2$ exhibits superconductivity after intercalating with water, suggesting intricate correlated physics inside the $CoO_2$ layers[9,10]. Previously, $Na_xCoO_2$ thin films have been deposited using pulsed-laser deposition (PLD)[11] on α-$Al_2O_3$ (001), $SrTiO_3$ (001), a bulk $Na_{3.4}Sc_{0.4}Zr_{1.6}(SiO_4)_2(PO_4)$ Nasicon solid electrolyte, and onto Au-coated MgO (001). These epitaxial $Na_xCoO_2$ model systems have been grown with their c-axis perpendicular to the plane of the substrate surface[12,13,10,9,14] (see Table 1). Yet films grown with the c-axis perpendicular to the substrate are disadvantageous for electrochemical intercalation because sodium-ion transport occurs along the layers of $CoO_2$ (normal to the c-axis). In a perfect crystal, the ions would need to migrate from the edges of the film to penetrate the material, which is prohibitively slow. Here, we synthesize an epitaxial thin film of $Na_xCoO_2$ with a c-axis canted to the substrate normal by combining molecular-beam epitaxy (MBE) and a topotactic transformation. We show that the films are epitaxial and demonstrate sodium extraction in an electrochemical cell (see Fig. 1).

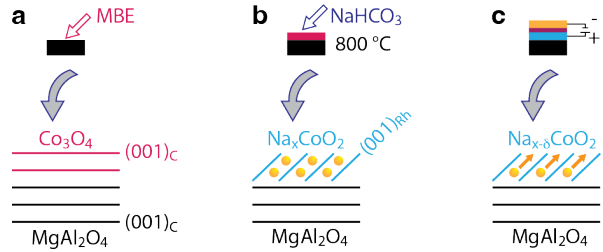

**Figure 1: Schematic of the study.** MBE-grown $Co_3O_4$ films (a) are topotactically transformed to $Na_xCoO_2$ (b), and subsequently electrochemically charged to $Na_{x-\delta}CoO_2$ (c).

| Research Group | Substrate (surface) | Film prior to topotactic transformation (surface) | Film after topotactic transformation/direct growth (surface) |
|---|---|---|---|
| Venimadhav et al. | α-$Al_2O_3$ (001) | $Co_3O_4$ (111) | Epitaxial $Na_xCoO_2$ (001) |
| Krockenberger et al. | α-$Al_2O_3$ (001) | N/A (direct growth) | Epitaxial $Na_xCoO_2$ (001) |
| Kehne et al. | $Na_{3.4}Sc_{0.4}Zr_{1.6}(SiO_4)_2(PO_4)$ Nasicon solid electrolyte | N/A (direct growth) | $Na_xCoO_2$ (001) |
| Shibata et al. | Au-deposited MgO (001) | N/A (direct growth) | Epitaxial $Na_xCoO_2$ (001) |
| Hildebrandt et al. | $SrTiO_3$ (001) | $Co_3O_4$ (111) | Epitaxial $Na_xCoO_2$ (001) |

**Table 1:** Methods of published $Na_xCoO_2$ thin film deposition all grown with their c-axis perpendicular to the plane of the substrate. [12,13,10,9,14] Hildebrandt et al. (Ref. [9]) also grew $Na_xCoO_2$ on $SrTiO_3$ (110), $SrTiO_3$ (111), c-cut $Al_2O_3$, and MgO (111), but results on the epitaxial orientation of those films and especially each film's c-axis orientation are limited.

**METHODS**

**Molecular-Beam Epitaxy**

Initial thin film growth was performed in the PARADIM (Platform for the Accelerated Realization, Analysis, and Discovery of Interface Materials) user facility. Chemo-mechanically polished 10 x 10 mm $MgAl_2O_4$ substrates (CrysTec GmbH) were used as received (the substrate surface is $(001)_C$, C: cubic). The $Co_3O_4$ thin films were synthesized in a Veeco Gen-10 MBE system at a substrate temperature of 350 °C and a background atmosphere of 10% $O_3$ + 90% $O_2$. The cell temperature used to deposit $Co_3O_4$ was 1238-1250 °C, depending on the cobalt flux from the effusion cell at the time of deposition. The typical flux used for $Co_3O_4$ deposition was $1\times10^{13}$ atoms/[cm$^2$s]. The quality of the $Co_3O_4$ was determined by examining the in-situ reflection high-energy electron diffraction (RHEED)[15,16] patterns during thin film deposition (see inset in Fig. 2a) and x-ray diffraction (XRD) after deposition (see Fig. 2a, top), where asterisks indicate substrate peaks and $Co_3O_4$ peaks, some of which overlap due to a small 0.01% lattice mismatch[17]. A water-lubricated diamond saw was used to cut each 10 x 10 mm film of $Co_3O_4$ deposited on a $MgAl_2O_4$ (001) substrate into 5 x 5 mm films before the topotactic transformation. Subsequent cleaning was done using water, followed by acetone and isopropyl alcohol (IPA).

**Topotactic transformation**

We topotactically transformed the films inside a pristine alumina single-bore furnace tube using the LabTemp-Tube-furnace-1700-C-1. The films were annealed in air in a bed of excess sodium bicarbonate and heated with a ramp rate of 300 °C/h. They were held at 800 °C for 2.3 h and ambiently cooled to room temperature to achieve the topotactic transformation[11]. The films were then cleaned again with IPA. After the topotactic transformation, the $Na_xCoO_2$ thin films became water-sensitive and degraded if exposed to water, as seen via x-ray diffraction (not shown). Therefore, we used no acetone or water to clean the films after the topotactic transformation and stored the samples inside a desiccator to prevent exposure to humidity in the air after the transformation step. Prior work by Venimadhav et al. used topotactic transformation of the $(111)_C$-oriented cubic structure utilizing sodium vapor[11]. The quality of $(111)_C$-oriented $Co_3O_4$ is inferior to $(001)_C$-oriented $Co_3O_4$ due to a higher energy surface that produces rougher films on sapphire by PLD and likely MBE. The $(001)_C$-oriented cubic $Co_3O_4$ had a smooth surface with a root-mean-square roughness of 50 pm as measured by atomic force microscopy (AFM, not shown) due to the low surface energy.

**Characterization of Thin Films**

X-ray diffraction (XRD) measurements were performed *ex-situ* after growth using PANalytical Empyrean and Rigaku SmartLab X-Ray diffractometers with Cu Kα-1 radiation[18]. AFM images are measured using Asylum Cypher Environmental AFM and Asylum MFP-3D-BIO AFM[19]. An Amscope 40X-500X Trinocular Dual-Illumination Upright Metallurgical Microscope with a 3MP attached camera and polarizers was used to acquire optical microscopy images[20]. A Zeiss Gemini 500 Scanning Electron Microscope was used to collect SEM images and EDS data (not shown)[21]. Thermo Fisher Spectra, cold-field emission Scanning Transmission Electron Microscope (STEM) was used [21,22]. The microscope was operated at 120 kV with a convergence angle of 30mrad. The final high-angle annular dark-field (HAADF) image was obtained by image registration of 20 fast (frame time = 0.5 s) images using Velox software.

**Electrochemical Measurements**

A three-electrode Swagelok setup was used to build a half-cell with pure sodium as the anode material and the thin film of $Na_xCoO_2$ on $MgAl_2O_4$ as the cathode material[23]. The sodium was stored in kerosene inside an Argon-filled glovebox to prevent oxidation before cell assembly. Half-cell assembly was done inside an Argon-filled glovebox. A holder for the half-cell was made using a 1/2" PFA Tee Compression Fitting, 316 Stainless Steel rods, a PEEK rod, and a conical compression spring. The holder was upright for all experiments using a small block of wood and zip ties. A Whatman Grade GF/D Glass Microfiber separator was used to separate active cathode material from active anode material in the half-cell. The electrolyte was made in-house. The solvent was 50 vol% ethylene carbonate and 50 vol% propylene carbonate[4]. The solute, $NaPF_6$, was then introduced in a 1 Mol concentration. The electrolyte was added to the half-cell in excess. Aluminum foil and stainless-steel mesh were used to wrap the thin film to allow for charge transport, given that the substrate used for thin film deposition had poor electrical conductivity. The aluminum foil and stainless-steel mesh were also used to limit edge effects. After charging and discharging, the film was removed from the cell and cleaned with IPA in preparation for characterization. We used Biologic's SP300

– a two-channel potentiostat/galvanostat with +/- 1 A (expandable up to 800 A)[24,25] – for electrochemical measurements.

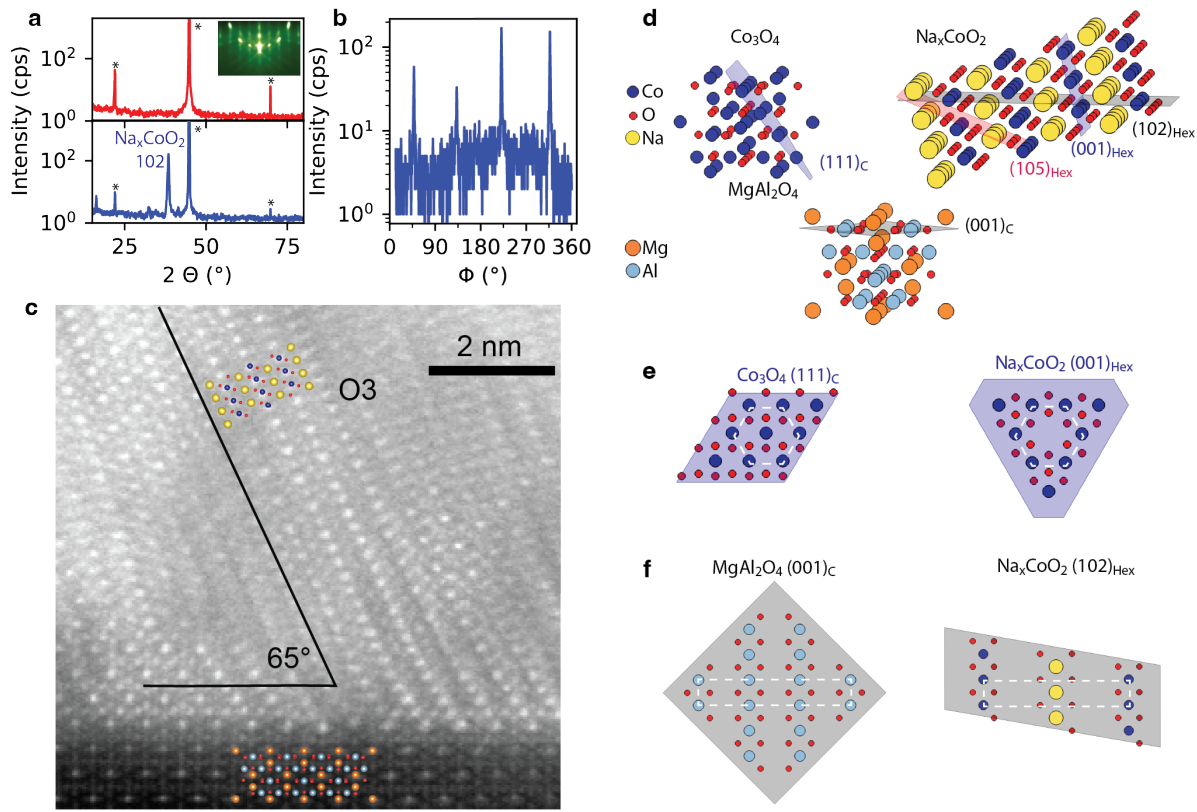

**Figure 2: Crystal structure characterization.** (a) XRD of $Co_3O_4$ deposited onto $MgAl_2O_4$ $(001)_C$ (top), and topotactically transformed $Na_xCoO_2$ on $MgAl_2O_4$ (bottom) measured along the $[001]_C$ direction. The strong substrate $004_C$ peak at $2\theta=44.8°$ has an intensity of $3\times10^5$ counts/s and is cut off for better visibility. (Inset) RHEED pattern of $Co_3O_4$ along $[110]_C$. (b) Azimuthal $\phi$-scan of $Na_xCoO_2$ $105_{Hex}$ peak measured with an inclination of $\chi=20°$ out of the scattering plane (see red plane in d). (c) STEM image of topotactically transformed $Na_xCoO_2$ on $MgAl_2O_4$, projected along $[1\bar{1}0]_C$, including the overlayed proposed O3-$Na_xCoO_2$ and $MgAl_2O_4$ crystal structures. (d) Proposed epitaxial relationship of $Co_3O_4$ on $MgAl_2O_4$ and $Na_xCoO_2$ on $MgAl_2O_4$ generated using VESTA [35]. (e) View of the $Co_3O_4$ $(111)_C$ and O3-$Na_xCoO_2$ $(001)_{Hex}$ planes (blue planes in d). The side-lengths of white dashed hexagons are 2.89 Å in $Ca_3O_4$ and 2.84 Å in O3-$Na_xCoO_2$. (f) View of the $MgAl_2O_4$ $(001)_C$ and O3-$Na_xCoO_2$ $(102)_{Hex}$ planes (gray planes in d). The side-lengths of white rectangles are (17.14 Å, 2.86 Å) in $MgAl_2O_4$ and (16.38 Å, 2.89 Å) in O3-$Na_xCoO_2$.

## RESULTS AND DISCUSSION

**Characterization of topotactically transformed structure**

The XRD data of topotactically transformed films show a new Bragg peak at $2\theta = 38.4°$, corresponding to a lattice spacing of about 2.4 Å (see Fig. 2a). To confirm that the transformation is topotactic and that the newly formed structure retains epitaxial alignment with the $MgAl_2O_4$ substrate, we collected an azimuthal XRD scan ($\varphi$-scan) of the film with the film normal rotated by $\chi = 21°$ out of the scattering plane, incident angle $\theta = 23°$, and scattering angle $2\theta = 46°$. The scan revealed four peaks approximately separated by 90° and rotated by 45° off the substrate edges, $[100]_C$ and $[010]_C$. The result confirmed epitaxy (the lack of the crystallographic registry with the substrate along one of its edges would generate no peaks in the φ-scan). STEM collected in projection along the $[\bar{1}10]_C$ direction is shown in Fig. 2c. The transformed structure has visible atomic layers and looks significantly different from the cubic structure in the substrate. The layers are inclined roughly 65° with respect to the film surface (see Fig. 2c).

**Proposed epitaxial alignment of $Na_xCoO_2$**

The proposed crystallographic orientation of $Na_xCoO_2$ is shown in Fig. 2d, together with the initial $Co_3O_4$ structure and the $MgAl_2O_4$ substrate. For the following analysis, we used $Co_3O_4$ (mp-18748, $Fd\bar{3}m$, No. 227), $MgAl_2O_4$ (mp-3536, $Fd\bar{3}m$, No. 227), and O3-$Na_xCoO_2$ (mp-18921, $R\bar{3}m$, No. 166) structures from materialsproject.org[26] and modified the lattice constants of O3-$Na_xCoO_2$ following Ref.[27] for x=1. The topotactic transformation process starts with $Co_3O_4$ $(001)_C$ plane oriented along the substrate surface $MgAl_2O_4$ $(001)_C$ (see Fig. 2d). The $Co_3O_4$ $(111)_C$ plane displays three-fold symmetry, which is also present in the $Na_xCoO_2$ $(001)_{Hex}$ plane (see Fig. 2e, Hex: hexagonal). Therefore, the $(001)_{Hex}$ $Na_xCoO_2$ plane likely replaces the $Co_3O_4$ $(111)_C$ plane: the mismatch between the Co-Co distances within these two planes is less than 2% (see Fig. 2e). We further hypothesize that after the transformation, the $(102)_{Hex}$ $Na_xCoO_2$ planes are parallel to the substrate surface. The calculated scattering angle of the $102_{Hex}$ peak is 37.7°, which is consistent with the measured peak at 38.4° (see Fig. 2a). In addition, the calculated angle between $(102)_{Hex}$ and $(001)_{Hex}$ of 72° is close to the 65° observed in STEM (see Fig. 2d). The discrepancy between experiment and model could be due to possible monoclinic distortions in $Na_xCoO_2$[27], lower sodium concentration resulting in lattice rearrangements[27], or substrate clamping. The proposed orientation of the $Na_xCoO_2$ with respect to the $MgAl_2O_4$ substrate is shown in Figure 2f. Within this model, the substrate imposes a small compressive strain within the layers of $Na_xCoO_2$ and a larger tensile strain in the direction that is almost perpendicular to the layers (see Fig. 2f). Finally, in the proposed orientation of $Na_xCoO_2$, the $105_{Hex}$ Bragg peak is accessible when $[110]_C$ lies within the scattering plane, and the film is rotated around $[110]_C$ by 20°, which is consistent with the results of our azimuthal $\phi$-scan. The calculated position of the $105_{Hex}$ Bragg peak, $2\theta$=46.5°, is in agreement with the measured value, $2\theta$=46°. Finally, because $Co_3O_4$ has four equivalent $\{111\}_C$ planes, any of these four can locally transform into the $Na_xCoO_2$ $(001)_{Hex}$ plane. As a result, four twins of the $Na_xCoO_2$ structure are present and visible as four $105_{Hex}$ reflections in the azimuthal $\phi$-scan (see Fig. 2b).

Our interpretation of the structural characterization data suggests a topotactic transformation into the O3-$Na_xCoO_2$ structure with $(102)_{Hex}$ plane parallel to the substrate interface (see Fig. 2) and the $(102)_{Hex}$ reflection having the largest intensity. Yet the XRD (Fig. 2a) also shows a much weaker peak at $2\theta = 16°$. This scattering angle corresponds to the $(001)_{Hex}$ lattice spacing and reveals $Na_xCoO_2$ with transition metal layers parallel to the substrate. Furthermore, some of the cubic $Co_3O_4$ structure appears to persist after the transformation (see the left bottom side of Fig. 2c and note the two peaks at $2\theta$=21.9° and $2\theta$=69.8° in Fig. 2a matching $002_C$ and $006_C$, which weaken about five-fold in intensity during the topotactic transformation yet remain present). Lei and coauthors reported the formation of different possible $Na_xCoO_2$ structures during solid-state synthesis at high temperatures and various Na-Co weight fractions: at intermediate sodium content, P2-$Na_xCoO_2$ is most stable due to entropy, and monoclinic distortions of P3 and O3 are possible[27]. Our XRD and STEM data cannot exclude the presence of these structures.

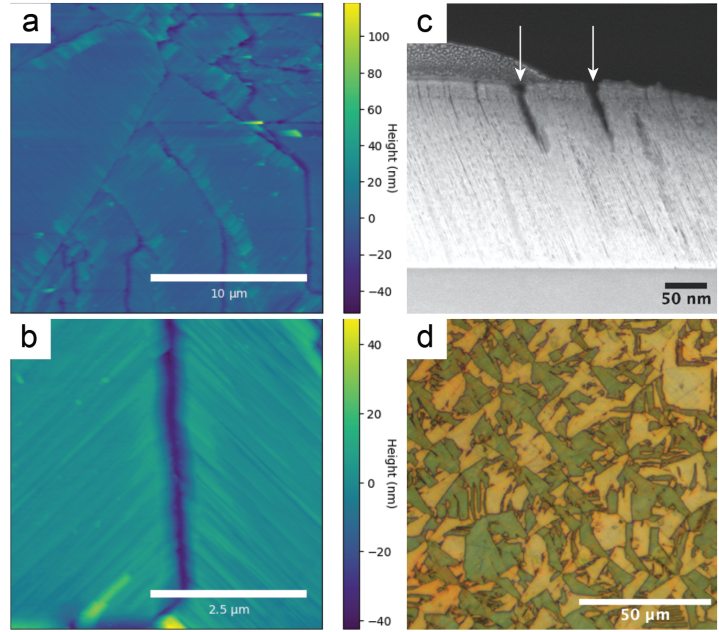

**Figure 3: Microstructure of topotactically transformed Na$_x$CoO$_2$.** (a,b) AFM of topotactically transformed Na$_x$CoO$_2$. (c) STEM image of Na$_x$CoO$_2$ on MgAl$_2$O$_4$ (001)$_C$ after charge and discharge. The attempts to image pristine Na$_x$CoO$_2$ were unsuccessful due to electron beam damage. The beam damage was much less severe in electrochemically cycled films. (d) Polarized light optical microscopy of Na$_x$CoO$_2$ after topotactic transformation.

**Microstructure of topotactically transformed Na$_x$CoO$_2$**

AFM images of the as-deposited Co$_3$O$_4$ films (not shown) reveal a smooth surface with a root-mean-square (RMS) height variation of 50 pm. After the topotactic transformation, the AFM data still indicates a flat surface, albeit with cracks visible as dark lines (see Fig. 3 a,b). The size of the largest crack-free domains is on the order of 10 μm x 10 μm. Within these crack-free domains, linear height modulations are visible, with lines running at approximately 45 ° with respect to the substrate edge [100]$_C$. We elucidate the nature of the lines visible on the surface with cross-sectional STEM (see Fig. 3c): the diagonal lines in Fig. 3 a,b are grooves visible in Fig. 3c (see white arrows) separated by about 100 nm. These groves are hypothesized to arise from delamination of the layered structure reported in similar layered oxides for sodium intercalation[28]. Only two types of lines are visible on the surface, consistent with the twinning of four {111}$_C$ planes into the (001)$_{Hex}$ plane; two twins produce the same orientation of groves on the surface. The optical microscope image taken with a polarizer and analyzer set at an angle of approximately 0 degrees from one another shows domains visible in AFM images. Polarized light microscopy can distinguish grain orientation in polycrystalline materials[29,30]. The contrast seen in this method is perhaps due to different crystallographic facets being exposed and creating anisotropic roughness, which reflects polarized light depending on orientation. Optical microscopy reveals a large variation in domain sizes. Overall, the topotactic transformation generates Na$_x$CoO$_2$ with large single crystalline domains, ~100 nm thick and hundreds of square microns in area, epitaxially aligned to the substrate.

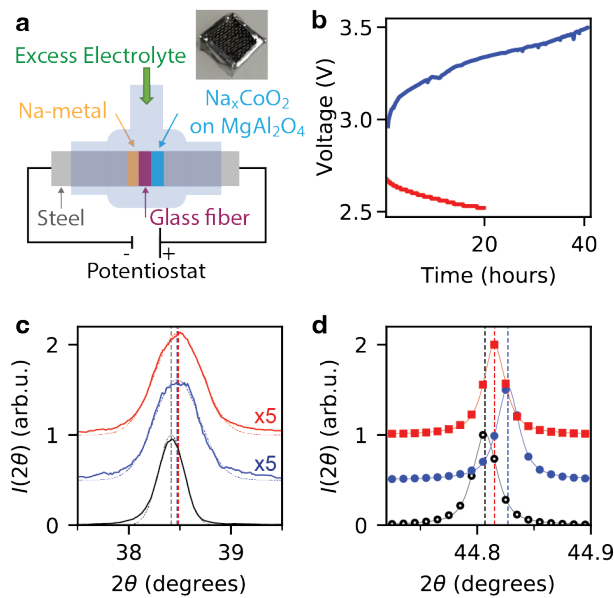

**Figure 4: Electrochemical cycling of Na$_x$CoO$_2$.** (a) A schematic of the electrochemical half-cell. (b) Voltage as a function of time during galvanostatic charge to 3.4 V (blue) and discharge to 2.5 V (red). (c,d) *Ex-situ* XRD of (c) Na$_x$CoO$_2$ 102$_{Hex}$ and (d) MgAl$_2$O$_4$ (004)$_C$ Bragg peak. The black/red/blue curves show data from pristine/charged/discharged specimen. In (c), solid lines (data) dashed lines (Gaussian fit) and in (d) symbols (data) and solid lines (pseudo-Voight fit) are shown. The vertical dashed lines in (c,d) show the center of the fit-function.

### Electrochemical cycling of topotactically transformed Na$_x$CoO$_2$

We charged (sodium extraction from Na$_x$CoO$_2$) and discharged (sodium insertion into Na$_x$CoO$_2$) the cell electrochemically (see Figs. 4a,b). During charge, the voltage increased monotonically from 3 V to 3.5 V, consistent with electrochemical sodium-ion extraction from Na$_x$CoO$_2$ in conventional nano-particulate morphology[8]. The discharge lasted significantly less time than the charge. Due to the air instability of the film, it is likely that the film self-discharged due to exposure to humidity during XRD data collection.

To confirm electrochemical sodium extraction, we conducted *ex-situ* XRD. After electrochemical charging of the cell, the XRD data showed a shift of the 102$_{Hex}$ peak to higher angles from $2\theta = 38.41°$ to $2\theta = 38.47°$, consistent with a shortening of the (102)$_{Hex}$ interplanar distance. When sodium exits the lattice upon charging, the (001)$_{Hex}$ lattice constant is expected to increase, and the (100)$_{Hex}$ lattice spacing is expected to decrease[27]. Using lattice parameters reported by Lei et al.[27], we calculated $2\theta = 37.74°$ for pristine O3-Na$_{1.00}$CoO$_2$ and $2\theta = 38.40°$ for charged P3-Na$_{0.56}$CoO$_2$. The Bragg peak shift, therefore, shows the expected tendency, albeit a smaller magnitude. Additionally, the intensity of the 102$_{Hex}$ Bragg peak reduced by a factor of 5, indicating a structural degradation of Na$_x$CoO$_2$ crystallinity during electrochemical charging (see Fig. 4c). After cathode discharge, the 102$_{Hex}$ peak did not shift back to its initial position: we observed no significant insertion of Na-ions into charged Na$_x$CoO$_2$. Possible reasons for the lack of electrochemical Na-ion insertion are structural degradation and surface reactions during charge. Another plausible explanation is the mechanical restraint of the thin film by the substrate: the (001)$_{Hex}$ lattice spacing is expected to shrink upon Na-ion insertion[27,28], yet our proposed epitaxial alignment suggests a large tensile strain oriented primarily perpendicular to the (001)$_{Hex}$ layers (see Fig. 2f) potentially preventing a decrease of the inter-layer spacing. We also observed a partially reversible shift in the MgAl$_2$O$_4$ substrate peak, from $2\theta = 44.807°$ to $2\theta = 44.827°$ during charge and $2\theta = 44.817°$ during discharge (see Fig. 4d). It is possible that electrochemical reactions also occurred in the MgAl$_2$O$_4$ substrate, which has been discussed as a material for multivalent Mg$^{2+}$ intercalation in aqueous solutions [31].

**CONCLUSIONS**

We used MBE to synthesize high-quality Co$_3$O$_4$ films and subsequently transformed the films to Na$_x$CoO$_2$. We presented evidence for the epitaxial registry of the film with respect to the substrate after the topotactic transformation. We also proposed a crystallographic mechanism for the phase transformation. By starting with high-quality (001)$_C$ cobalt oxide, rather than the conventionally oriented (111)$_C$ cobalt oxide, we synthesized a uniquely oriented film of Na$_x$CoO$_2$ with CoO$_2$ layers having a large inclination with the film surface. Using an electrochemical cell, we extracted sodium ions from epitaxial Na$_x$CoO$_2$, as confirmed by *ex-situ* XRD. In future, the option of substituting Co with other transition metals (*M*) during the atomic deposition of spinel oxides *M*$_3$O$_4$ potentially provides a novel tool for the controlled synthesis of Na$_x$*M*O$_2$ with various combinations of transition metals. The system thus represents a model system for further characterization of layered oxides for sodium-ion intercalation by using a multimodal approach demonstrated here. We anticipate an expansion to operando measurements to be straightforward and future *operando* XRD to allow studying subtle crystal rearrangements such as Jahn-Teller distortions[32], operando x-ray reflectivity to reveal surface reactions at the solid-electrolyte-interface[33], operando optical microscopy, and operando x-ray nanoimaging to visualize the microstructure[34]. As such, this epitaxial model system will likely advance our understanding of alkaline-ion intercalation into layered oxides for energy storage.

**ACKNOWLEDGMENTS**

This research was primarily supported by the National Science Foundation under Grant No. CAREER DMR 1944907 (S.D.M., J.H, A.S: topotactic transformations, electrochemical testing, optical microscopy, atomic force microscopy, x-ray characterization). This work made use of the Platform for the Accelerated Realization, Analysis, and Discovery of Interface Materials (PARADIM) (J.S., D.G.S., thin film synthesis, film characterization) supported by the National Science Foundation under Cooperative Agreement No. DMR-2039380. The authors thank Regina Garcia-Mendez for her help in designing a three-electrode experiment at Cornell Energy Systems Institute (CESI) and Hanjong Paik for assistance in the use of the PARADIM facility. Luka Radosavljevic assisted in AFM measurements, Phil Carubia trained on how to use the annealing furnace, and Mark Pfeifer assisted in the collection of the azimuthal scan. Thanks to Jakob Gollwitzer for helping with additional AFM measurements. Thanks to Mick Thomas for help with SEM, and John L. Grazul for help with optical microscopy. Thanks to Noah Schnitzer for helping to confirm the overlay of our expected structure onto our STEM image. Electron microscopy work was supported by the National Science Foundation (Platform for the Accelerated Realization, Analysis, and Discovery of Interface Materials (PARADIM)), under Cooperative Agreement No. DMR-2039380. This work made use of a Helios FIB supported by NSF (No. DMR-1539918) and the Cornell Center for Materials Research Shared Facilities, which are supported through the NSF MRSEC program (No. DMR-1719875). Substrate preparation was performed in part at the Cornell NanoScale Facility, a member of the National Nanotechnology Coordinated Infrastructure, which is supported by the NSF (Grant No. NNCI-2025233). The authors thank Sean C. Palmer for his assistance with substrate preparation.